\documentclass[a4paper]{article}

\usepackage{amsmath, amssymb}
\usepackage{graphicx, subfigure}

\newcommand{\keywords}[1]{{\bf Keywords:} {#1}}
\newcommand{\subclass}[1]{{\bf Mathematics Subject Classification (2000):} {#1}}
\newcommand{\JEL}[1]{{\bf JEL:} {#1}}

\newcommand{\MakeTitle}{\maketitle\newcommand{\and}{$\cdot$ }}

\title{A Note on ``A Family of Maximum Entropy Densities Matching Call Option Prices''}

\author{
	Cassio Neri
	\thanks{Lloyds Banking Group, \texttt{cassio.neri@lloydsbanking.com}.}
	\and
	Lorenz Schneider
	\thanks{Center for Financial Risks Analysis (CEFRA), EMLYON Business School, \texttt{schneider@em-lyon.com}.}
}

\date{December 18, 2012}

\begin{document}

\MakeTitle

\begin{abstract}

In \cite{NeriSchneider2009} we presented a method to recover the Maximum Entropy Density (MED) inferred from prices of call and digital options on a set of $n$ strikes.
To find the MED we need to numerically invert a one-dimensional function for $n$ values and a Newton-Raphson method is suggested.
In this note we revisit this inversion problem and show that it can be rewritten in terms of the Langevin function for which numerical approximations of its inverse are known.

The approach of \cite{NeriSchneider2009} is very similar to that of Buchen and Kelly (BK) \cite{BuchenKelly1996} with the difference that BK only requires call option prices.
Then, in \cite{NeriSchneider2011} we presented another approach which uses call prices only and recovers the same density as BK with a few advantages, notably, numerical stability.
\cite{NeriSchneider2011} provides a detailed analysis of convergence and, in particular, gives various estimates of how far (in different senses) the iterative algorithm is from the solution.
These estimates rely on a constant $m>0$.
The larger $m$ is the better the estimates will be.
A concrete value of $m$ is suggested in \cite{NeriSchneider2011}, and this note provides a sharper value.

\bigskip

\keywords{Entropy \and Information Theory \and $I$-Divergence \and Asset Distribution \and Option Pricing}

\bigskip

\subclass{91B24 \and 91B28 \and 91B70 \and 94A17}

\bigskip

\JEL{C16 \and C63 \and G13}
\end{abstract}

\section{Introduction}

In \cite{NeriSchneider2009} we presented a method to obtain the Maximum Entropy Density (MED) inferred from arbitrage free prices of call and digital options.
More precisely, the option prices are assumed for a given maturity and a set of $n$ strikes.
A similar approach was firstly derived in \cite{BuchenKelly1996} by Buchen and Kelly (BK) but they used call prices only.

The inclusion of digital prices simplified the methodology by requiring the solution of $n$ one-dimensional non-linear equations rather than the solution of BK's $n$-dimensional non-linear problem.
The numerical method that we presented turned out to be more stable than BK's.
Actually, Buchen and Kelly themselves mentioned the instability in their article.
Orozco Rodriguez and Santosa \cite{RodriguezSantosa2012} have also clearly pointed that out.

In general, when digital prices are prescribed, the MED obtained does not match BK unless the digital prices are the same as those implied by the BK density.
The algorithm presented in our second work \cite{NeriSchneider2011} does not assume that the digital prices are given and recovers the BK density without the instability issues.

The key point to recovering the BK density is to find the right set of digital prices.
This set is characterized by two equivalent conditions:
\begin{itemize}
\item It is the set of digital prices that produces a continuous density.
\item It is the set of digital prices which implies the density with highest entropy.
\end{itemize}
The latter condition suggests a clear way to recover the BK density by solving an optimization problem which turns out to be more stable than BK's original one.
In addition, the $n\times n$ Hessian matrix of the objective function of our approach is tridiagonal in contrast to the full matrix that appears in BK's algorithm.

Although our first algorithm does not generally produce the BK density, it is still needed because it is run at each step of our second algorithm.
In particular, we still need to solve the $n$ one-dimensional problems mentioned above.
We turn our attention back to this problem.

In the proof of Proposition 2.3 of \cite{NeriSchneider2009} we have shown that, through a simple change of variables, solving the $n$ equations is equivalent to inverting a single function at $n$ different points.
We noticed that this function has the profile of a probability distribution function 
(it is increasing and has limits $0$ and $1$ when the argument goes to $-\infty$ and $+\infty$, respectively) 
and we wondered if an approximation of this function's inverse was already known.
In this note we shall see that this is indeed the case.

The function to be inverted is related to the Langevin function.
This function also allows us to improve some results regarding the convergence analysis of our algorithm presented in \cite{NeriSchneider2011}.

\section{Finding the MED through the Langevin Function}

Let $K_0 := 0 < K_1 < ... < K_n < K_{n+1} := \infty$.
As in \cite{NeriSchneider2011}, for $i=0,...,n$ we define
\[
c_i(\beta) := \left\{
\begin{array}{cl}
\displaystyle \ln \left( \frac{e^{\beta K_{i+1}} - e^{\beta K_i}}{\beta} \right) \quad & \text{for } i < n \text{ and } \beta \neq 0, \\
\\
\displaystyle \ln(K_{i+1} - K_i) & \text{for } i < n \text{ and } \beta = 0,\\
\\
\displaystyle \ln\left(-\frac{e^{\beta K_i}}{\beta}\right) & \text{for } i = n \text{ and } \beta < 0,
\end{array}
\right.
\]

The function $c_n$ has a different treatment and, as seen in \cite{NeriSchneider2009}, its derivative is easily inverted.
Therefore, we focus on the case $i < n$.

Recall that the apparent singularity at $\beta=0$ is not effective and $c_i$ is twice continuously differentiable.

For each $i<n$, let $U_i := \frac12(K_{i+1}+K_i)$ and $V_i := \frac12(K_{i+1}-K_i)$.
For $i<n$ and $\beta\ne0$ we have
\begin{align*}
c_i(\beta)
&= \ln \left( \frac{e^{\beta K_{i+1}} - e^{\beta K_i}}{\beta} \right)
 = \ln \left( e^{\beta (U_i + V_i)} - e^{\beta (U_i - V_i)} \right) - \ln \beta \\
&= \beta U_i + \ln \left( 2\left(\frac{e^{\beta V_i} - e^{-\beta V_i}}{2}\right) \right) - \ln \beta
 = \beta U_i + \ln 2 + \ln \left( \sinh(\beta V_i) \right) - \ln \beta.
\end{align*}

\begin{figure}[ht]
\centering
\includegraphics[width=.6\textwidth]{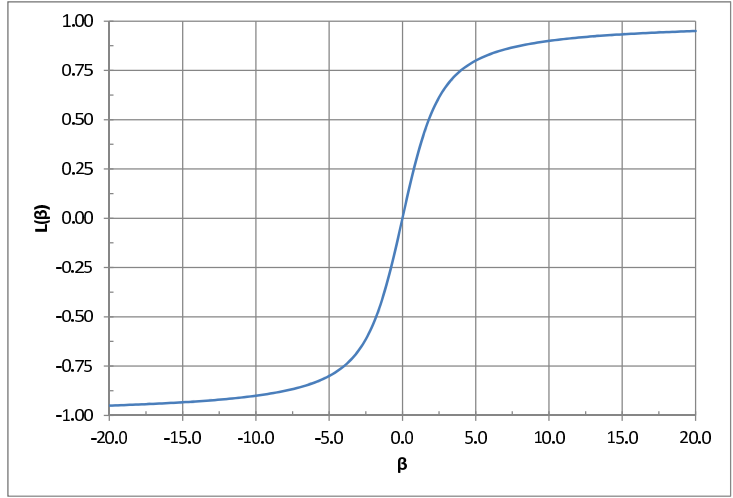}
\caption{Graph of the Langevin function.}
\label{Fig:1}
\end{figure}

Therefore,
\begin{equation}
\label{Eq:c'}
c_i'(\beta)
 = U_i + V_i\ \frac {\cosh(\beta V_i)}{\sinh(\beta V_i)} - \frac{1}{\beta}
 = U_i + V_i \left(\coth(\beta V_i) - \frac{1}{\beta V_i}\right)
 = U_i + V_i L(\beta V_i),
\end{equation}
where $L(x) := \coth(x) - x^{-1}$ is the Langevin function.
The graph of $L$ is shown in Figure \ref{Fig:1}.

Recall that finding the MED, given call and digital prices, requires solving for $\beta_i$
\[
c_i'(\beta_i) = \bar{K_i}
\]
where $\bar{K_i}\in (K_i, K_{i+1})$ is given in terms of digital and call prices only.
Hence, we need to solve
\[
U_i + V_iL(\beta V_i) = \bar{K_i}
\]
which gives
\[
\beta_i = \frac1{V_i} L^{-1}\left(\frac{\bar{K_i} - U_i}{V_i}\right).
\]

A few approximation formulas for $L^{-1}$ have been proposed in the literature.
Firstly, through the Taylor expansion around $x=0$,
\[
L^{-1}(x) = 3x + \frac{9}{5}x^3 + \frac{297}{175}x^5 + \frac{1539}{875}x^7 + ...
\]
Notice that the domain of $L^{-1}$ is $(-1, 1)$ and $L^{-1}(x)\rightarrow\pm\infty$ when $x\rightarrow\pm 1$.
Therefore a growing number of terms of the Taylor expansion is needed when $|x|\rightarrow1$.
To overcome this issue, from the Taylor expansion above one can derive the following Pad\'e approximant  \cite{Cohen1991}
\[
L^{-1}(x) = x \frac{3 - \frac{36}{35}x^2}{1-\frac{33}{35}x^2} + {\cal O}(x^6).
\]
However, the poles of this approximation are located at $x=\pm\sqrt{35/33}$ and not at $\pm1$ as those of $L^{-1}$.
Finally, rounding $36/35 \approx 1$ and $33/35\approx 1$ yields the rounded Pad\'e approximant \cite{Cohen1991}
\[
L^{-1}(x) = x \frac{3 - x^2}{1 - x^2}.
\]

Another approximation formula with a relative error smaller than $6.4\cdot 10^{-4}$ \cite{Bergstrom1999} is given by
\[
L^{-1}(x) = \left\{
\begin{aligned}
& 1.31446\cdot\tan(1.58986x) + 0.91209 x, & \quad & \text{if }|x| < 0.84136, \\
& 1 / (\mathrm{sign}(x) - x),             & \quad & \text{if } 0.84136 < |x| < 1.
\end{aligned}
\right.
\]

\section{Convergence Analysis}

We recall some notation from \cite{NeriSchneider2011} before stating stronger results for the convergence analysis presented in Section 6.2 of that article.

For a given maturity, we fix a set of undiscounted arbitrage-free call prices $\tilde{C}_1$, ..., $\tilde{C}_n$ at prescribed strikes $K_1$, ..., $K_n$ and also the forward asset price $\tilde{C}_0$ (which is seen as the price of a call option with strike $K_0 := 0$).
These prices impose a number of constraints on the values that undiscounted arbitrage-free digital prices $\tilde{D}_1$, ..., $\tilde{D}_n$ on the same strikes can assume.
Let $\Omega$ be the set of $\tilde D :=(\tilde{D}_1, ..., \tilde{D}_n)$ that verify these constraints.

For each $\tilde{D}\in\Omega$ the results of \cite{NeriSchneider2009} show that there exists a density $g_{\tilde{D}}$ which maximises entropy subject to implying the same prices of calls $\tilde{C}_0$, $\tilde{C}_1$, ..., $\tilde{C}_n$ and digitals $\tilde{D}_1$, ..., $\tilde{D}_n$.
Let $H(\tilde{D})$ be the entropy of such a density.
By varying $\tilde{D}$ in order to maximise $H$, we recover the BK density $g_{\hat{D}}$ and its corresponding vector of digital prices $\hat{D}$.

Proposition 6.1 of \cite{NeriSchneider2011} introduces a constant $m>0$ that drives the subsequent convergence analysis.
More precisely, it is shown that for all $\tilde{D}\in\Omega$ we have
\begin{enumerate}
\item $H(\hat{D}) - H(\tilde D) \le \frac1{2m}\|H'(\tilde{D})\|^2$;
\item $\|\hat{D}-\tilde{D}\| \le \frac2m \|H'(\tilde{D})\|$; and
\item $\|g_{\hat{D}} - g_{\tilde{D}}\|_{L^1} \le \frac1{\sqrt{m}} \|H'(\tilde{D})\|$.
\end{enumerate}

From the relations above, we see that the bigger $m$ is, the better our knowledge of the convergence becomes.
In other words, the number of steps required to get an approximated solution within a given tolerance decreases.
Proposition 6.1 also suggests taking $m = 4\sin^2(\pi/(2n+2))$, but this value approaches $0$ as the number of strikes goes to infinity.
In the following, we shall see that we can improve this result and take $m = 4\sin^2(\pi/(2n+2)) + 1/2$.

Looking at the proof of Proposition 6.1, we see that $m$ is given by the sum of $m_1 := 4\sin^2(\pi/(2n+2))$ and
\[
m_2 := \frac12
\min\left\{
\frac{(K_1 - \bar{K}_0)^2}{p_0c_0''(\beta_0)}, \quad
\min_{i = 1, ..., n-1}\left\{
\frac{(\bar{K}_i - K_i)^2}{p_ic_i''(\beta_i)}, \quad
\frac{(K_{i+1} - \bar{K}_i)^2}{p_ic_i''(\beta_i)}
\right\}, \quad
\frac{(\bar{K}_n - K_n)^2}{p_nc_n''(\beta_n)}
\right\},
\]
where $p_i = \tilde{D}_i - \tilde{D}_{i+1}$ and $\bar{K}_i = c_i'(\beta_i)$ for all $i = 0, ..., n$.
We shall see that $m_2 \ge 1/2$ and for this, it suffices to show that
\[
\frac{(c_i'(\beta_i) - K_i)^2}{p_ic_i''(\beta_i)} \ge 1 \quad \forall i=1, ..., n,
\]
and
\[
\frac{(K_{i+1} - c_i'(\beta_i))^2}{p_ic_i''(\beta_i)} \ge 1 \quad \forall i=0, ..., n-1.
\]
The arguments are similar and we shall show only the first relation above.

Firstly, we notice that $p_i\le 1$; therefore, it is enough to show that
\[
\frac{(c_i'(\beta_i) - K_i)^2}{c_i''(\beta_i)} \ge 1 \quad \forall i=1, ..., n.
\]
The case $i=n$ is straightforward and is left to the reader.
In the sequel we assume $i<n$.

Using \eqref{Eq:c'} and $U_i - K_i = V_i$ yields
\[
(c_i'(\beta_i) - K_i)^2 = (U_i + V_i L(\beta_i V_i) - K_i)^2 = V_i^2 (1 + L(\beta_i V_i))^2
\]
and
\[
c_i''(\beta_i) = V_i^2L'(\beta_i V_i).
\]
Hence,
\[
\frac{(c_i'(\beta_i) - K_i)^2}{c_i''(\beta_i)} = \frac{(1 + L(\beta_i V_i))^2}{L'(\beta_i V_i)}.
\]
To finish the proof we shall show that
\[
(1 + L(x))^2 \ge L'(x) \quad \forall x\in\mathbb{R}.
\]
For $x=0$ the result follows from $L(0)=0$ and $L'(0)=1/3$.
For $x\ne 0$, the result is equivalent to
\begin{align*}
\left(1 + \coth x - x^{-1}\right)^2 \ge 1 - \coth^2 x + x^{-2}
&\Longleftrightarrow \coth^2 x + 2\coth x - 2x^{-1} - 2x^{-1}\coth x \ge - \coth^2 x \\
&\Longleftrightarrow \coth^2 x + \coth x - x^{-1} - x^{-1}\coth x \ge 0 \\
&\Longleftrightarrow \left(\coth x - x^{-1}\right)\left(\coth x + 1\right) \ge 0 \\
&\Longleftrightarrow L(x)\left(\coth x + 1\right) \ge 0.
\end{align*}
The last inequality follows from the fact that $L(x)$ and $\coth x + 1$ both have the same sign as $x$.

\bibliographystyle{plain}
\bibliography{../bibtex/articles,../bibtex/books}

\end{document}